\pdfoutput=1

\documentclass[12pt]{nature}
\usepackage{iclr2019_conference,times}

\usepackage{graphics}
\usepackage{graphicx}
\usepackage{amsmath}

\usepackage[inline]{enumitem}
\title{Augmenting the Pathology Lab: An Intelligent Whole Slide Image Classification System for the Real World}

\author{Julianna D. Ianni$^{*,1}$, Rajath E. Soans$^{*,1}$, Sivaramakrishnan Sankarapandian$^1$, Ramachandra Vikas Chamarthi$^1$, Devi Ayyagari$^1$, Thomas G. Olsen$^{2,3}$, Michael J. Bonham$^1$, Coleman C. Stavish$^1$, Kiran Motaparthi$^4$, Clay J. Cockerell$^5$, Theresa A. Feeser$^1$, Jason B. Lee$^6$}

\begin{document}

\maketitle
\\
\\$*$These authors contributed equally to this work.
\begin{affiliations}
 \item Proscia Inc., Philadelphia, Pennsylvania, USA.
 \item Department of Dermatology, Boonshoft School of Medicine, Wright State University School of Medicine, Dayton, Ohio, USA
 \item Dermatopathology Laboratory of Central States, Dayton, Ohio, USA
 \item Department of Dermatology, University of Florida College of Medicine, Gainesville, Florida
 \item Cockerell Dermatopathology, Dallas, Texas, USA.
 \item Departments of Dermatology and Cutaneous Biology, Sidney Kimmel Medical College at Thomas Jefferson University, Philadelphia, Pennsylvania, USA.
\end{affiliations}
\section*{Abstract}
Standard of care diagnostic procedure for suspected skin cancer is microscopic examination of hematoxylin \& eosin stained tissue by a pathologist. Areas of high inter-pathologist discordance 
and rising biopsy rates
necessitate higher efficiency and diagnostic reproducibility.
We present and validate a deep learning system which classifies digitized dermatopathology slides into 4 categories. 

The system is developed using 5,070 images from a single lab, and tested on an uncurated set of 13,537 images from 3 test labs, using whole slide scanners manufactured by 3 different vendors. 
The system's use of deep-learning-based confidence scoring as a criterion to consider the result as accurate yields an accuracy of up to 98\%, and makes it adoptable in a real-world setting.
Without confidence scoring, the system achieved an accuracy of 78\%.
We anticipate that our deep learning system will serve as a foundation enabling faster diagnosis of skin cancer, identification of cases for specialist review, and targeted diagnostic classifications.

\section{Introduction}

Every year in the United States, 12 million skin lesions are biopsied,\cite{pathology_market} with over 5 million new skin cancer cases diagnosed.\cite{nmscIncidence}
After a skin lesion is biopsied, the tissue is fixed, embedded, sectioned, and stained with hematoxylin and eosin (H\&E) on glass slides, ultimately to be examined under microscope by a dermatologist, general pathologist or dermatopathologist who provides a diagnosis for each tissue specimen. Owing to the large variety of over 500 distinct skin pathologies\cite{feramisco2009phenotypic} and the severe consequences of a critical misdiagnosis,\cite{olhoffer2002histopathologic} diagnosis in dermatopathology demands specialized training and education.
Although the inter-observer concordance rate in dermatopathology is estimated to be between 90 and 95\%,\cite{Kent,Shah} there are some distinctions which present frequent disagreement among pathologists, such as in the case of melanoma vs. melanocytic nevi.\cite{farmer1996discordance, concordance1, concordance2,concordance3,Shoo}
Any system which could improve diagnostic accuracy provides obvious benefits for dermatopathology labs and patients; however, there are substantial benefits also to improving the distribution of pathologists' workloads.\cite{baidoshvili2018evaluating, ho2014can, hanna2019whole}  This can reduce diagnostic turnaround times in several scenarios. For example, when skin biopsies are interpreted initially by a dermatologist or a general pathologist, prior to referral to a dermatopathologist, it can result in a delay of days, sometimes in critical cases.
In another common scenario, additional staining is required to identify characteristics of the tissue not captured by standard H\&E staining. If those additional stains are not ordered early enough, there can be further delays to diagnosis. An intelligent system to distribute pathology workloads could alleviate some of these bottlenecks in lab workflows.
The rise in adoption of digital pathology\cite{pathology_market,aljanabi} provides an opportunity for the use of deep learning-based methods for closing these gaps in diagnostic reliability and efficiency.\cite{cruz2017accurate, litjens2016deep}
\par In recent years, deep neural networks have proven capable of identifying diagnostically relevant patterns in radiology and pathology.\cite{lungnature,Olsen,dermlevel,Li2019,Abramoff,enlitic,hwang,campanella}
While deep learning applied to medical imaging-based diagnostic applications has progressed beyond proof-of-concept,\cite{dermlevel, enlitic, Abramoff, hwang,lungnature} the translation of these methods to digital pathology must overcome unique challenges. Among these is sheer image size; a whole slide image (WSI) can contain several gigabytes of data and billions of pixels. Additionally, non-standardized image appearance (variability in tissue preparation, staining, scanned appearance, presence of artifacts) and a large number of pathologic abnormalities that can be observed present unique barriers to development of deployable deep learning applications in pathology. For example, Tellez et al.\cite{Tellez2019} demonstrate the strong impact that inter-site variance-- with respect to stain and other image properties-- can have on deep learning models.
Nonetheless, deep learning-based methods have recently shown promise in a number of tasks in digital pathology, primarily for segmentation models and networks which classify small patches within a WSI.\cite{Olsen,Korbar,sornapudi2018deep ,breastcamelyon,Ciompi2017,Bulten2019,Ghaznavi2013,bejnordi2018using,Janowczyk2016,Tellez2019,Hart} More recent methods have performed direct WSI classification.\cite{campanella,Ing2018,Li2019} However, many focus only on a single diagnostic class to make binary classifications,\cite{Olsen, campanella, Ing2018, Bulten2019} the utility of which breaks down in addressing subspecialties for which there is more than one relevant pathology of interest. Additionally, many of these methods have focused on curated datasets consisting of fewer than 5 pathologies with little diagnostic and image variability.\cite{Li2019,Olsen, Ing2018,Bulten2019}
\par The insufficiency of models developed and tested using small curated datasets such as CAMELYON\cite{breastcamelyon} was effectively demonstrated by Campanella et. al.\cite{campanella} However, while this study claimed to validate on data free of curation, the data presented featured limited capture of not only biological variability (e.g. exclusion of commonly-occurring prostatic interaepithelial neoplasia and atypical glandular morphologies) but also image variability originating from slide preparation and scanning characteristics (e.g. exclusion of slides with pen markings, need for retrospective human correction of select results, and poorer performance on externally-scanned images).
In contrast to deep learning systems exposed to contrived pathology problems and datasets, pathologists are trained to recognize hundreds of morphological variants of diseases they are likely to encounter in their careers and must adapt to variations in tissue preparation and staining protocols.
In addition to these variations, deep learning algorithms can also be sensitive to image artifacts. Some research has attempted to account for these issues by detecting and pre-screening image artifacts, either by automatically\cite{Kohlberger2019, DeepFocus,histoqc} or manually removing slides with artifacts.\cite{Bulten2019,Olsen,campanella} Campanella et. al\cite{campanella} include variability in allowed artifacts which others lack, but still selectively exclude images with ink markings, which have been shown to affect predictions of neural networks.\cite{Ali2019}

A real-world deep learning pathology system must be demonstrably robust to these variations. It must be tested on non-selected specimens, with no exclusions and no manual pre-screening of slides input or post-screening of the system outputs. A comprehensive test set for robustly assessing system performance should contain images: 
\begin{enumerate}
    \item From multiple labs, with markedly varied stain and image appearance due to imaging using different whole slide image scanner models and vendors, and variability in tissue preparation and staining protocols.
    \item Wholly representative of a diagnostic workload in the subspecialty (i.e. not excluding pathologic or morphologic variations which occur in a sampled time-period).
    \item With a host of naturally-occurring and human-induced artifacts: scratches, tissue fixation artifacts, air bubbles, dust and dirt, smudges, out-of-focus or blurred regions, scanner-induced misregistrations, striping, pen ink or letters on slides, inked tissue margins, patching errors, noise, color/calibration/light variations, knife-edge artifacts, tissue folds, and lack of tissue present.
    \item With no visible pathology (in some instances), or with no conclusive diagnosis, covering the breadth of cases occurring in diagnostic practice. 
\end{enumerate}
In this work, we present a pathology deep learning system (PDLS) which is capable of classifying WSIs containing H\&E-stained skin biopsies or excisions into diagnostically-relevant classes (Basaloid, Squamous, Melanocytic and Other). A key aspect of our system is that it returns a measure of confidence in its assessment; this is necessary in such classifications because of the wide range of variability in the images. A real-world system should not only return accurate predictions for commonly occurring diagnostic entities and image appearances, but also flag the non-negligible remainder of images whose unusual features lie outside the range allowing reliable model prediction. The PDLS is developed using 5,070 WSIs from a single lab (\emph{"Reference Lab"}), and independently tested on a completely uncurated and unrefined set of 13,537 sequentially accessioned H\&E-stained images from 3 additional labs, each using a different scanner and different staining and preparation protocol. No images were excluded. To our knowledge, this test set is the largest in pathology to date.
Our PDLS satisfies all the criteria listed above for real-world assessment, and is therefore to our knowledge the first truly real-world-validated deep learning system in pathology.

\section{Results}

\subsection{Overview and Evaluation of PDLS}
\begin{figure}
    \centering
    \includegraphics[width = \linewidth]{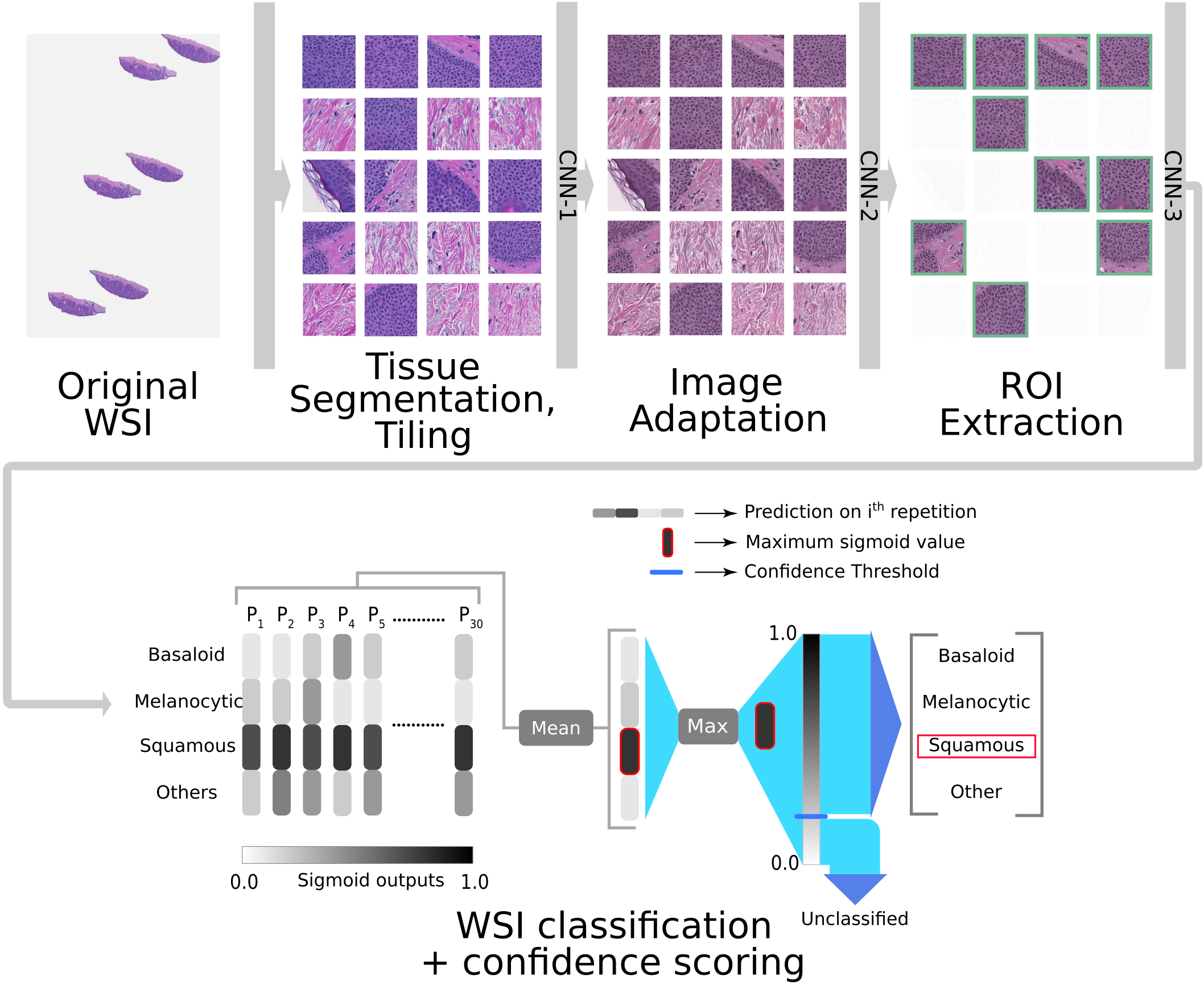}
    \caption{The process of classifying a whole slide image (WSI) with the pathology deep learning system is shown. The input WSI is first segmented and divided into tissue patches (Tissue Segmentation, Tiling); those patches pass through CNN-1, which adapts their stain and appearance to the target domain; they then pass through CNN-2 which identifies the regions of interest (patches) required to pass to CNN-3, which performs a 4-way classification, and repeats this 30 times to yield 30 predictions, where each prediction $P_i$ is a vector of dimension $N_{classes}$=4; the max of the class-wise mean of sigmoid output is the confidence score. If the confidence score surpasses a pre-defined threshold, the corresponding class decision is assigned.}
    \label{fig:processFig}
\end{figure}

The proposed system, as illustrated in Fig. \ref{fig:processFig}, takes as input a WSI and classifies it using a cascade of three independently-trained convolutional neural networks (CNNs) as follows: The first (\emph{CNN-1}) adapts the image appearance to a common feature domain, accounting for variations in stain and appearance; the second (\emph{CNN-2}) identifies regions of interest (ROI) for processing by the final network (\emph{CNN-3}), which classifies the WSI into one of 4 classes defined broadly by their histologic characteristics---Basaloid, Melanocytic, Squamous, and Other, as further described in Methods. Although the classifier operates at the level of an individual WSI, some specimens are represented by multiple WSIs, and therefore these predictions are aggregated to produce a single specimen-level classification. The classifier is trained such that for each image a predicted class is returned along with a confidence in the accuracy of the outcome. This allows discarding of predictions that are determined by the PDLS as likely to be false.

Since there is a large amount of variation in both pathologic findings of skin lesions as well as scanner or preparation-induced abnormalities, it is very important for the model to assess a confidence score for each decision; thereby, likely-misclassified images can be flagged as such. We developed a method of confidence scoring based on Gal et al.\cite{gal2016dropout} and set confidence thresholds a priori based only on performance on the validation set of the Reference Lab, which is independent of the data for which we report all measures of system performance (see Methods).
Three confidence thresholds were calculated and fixed based on the Reference Lab validation set such that discarding specimens with lower scores achieved the following 3 levels of accuracy in the remainder: 90\% (Level 1), 95\% (Level 2) and 98\% (Level 3).

\begin{figure}
    \centering
    \includegraphics[width = \linewidth]{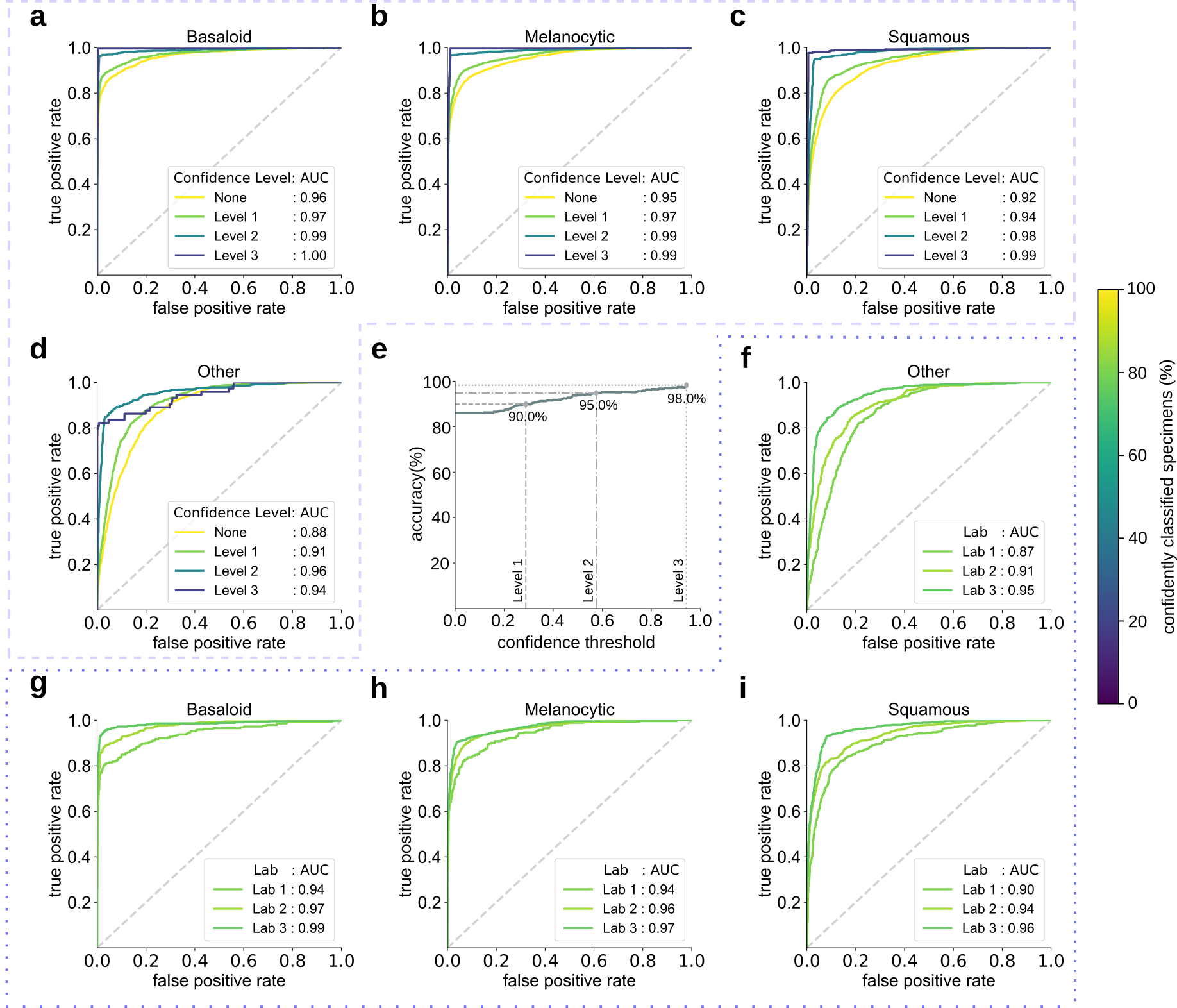}
    \caption{Receiver operating characteristic (ROC) curves are shown by lab, class, and confidence level for the test set of 13,537 images. ROC curves are shown for Basaloid (a,g), Melanocytic (b,h), Squamous (c,i) and Other (d,f) classes, with percentage of  specimens classified for each curve represented by the color bar at right. The three curves in each of (a-d) represent the respective thresholded confidence levels or no confidence threshold ("None"). The three curves in each of (f-i) represent the three labs. (e) Validation set accuracy in the Reference Lab is plotted versus sigmoid confidence score, with dashed lines corresponding to the sigmoid confidence thresholds set (and fixed) at 90\% (Level 1), 95\% (Level 2), and 98\% (Level 3). }
    \label{fig:ROC}
\end{figure}

\begin{figure}
     \centering
     \includegraphics[width = \linewidth]{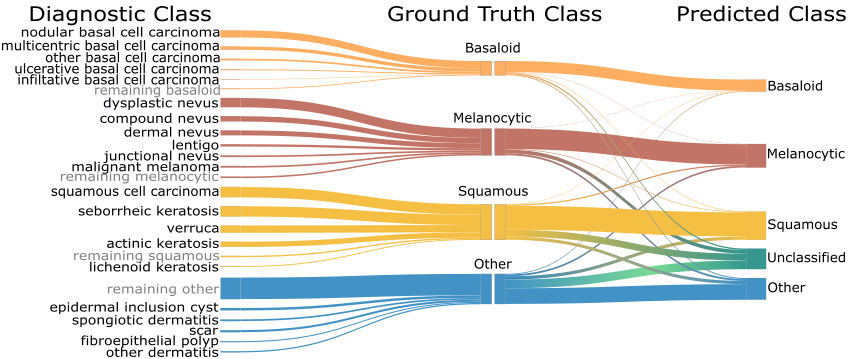}
     \caption{Sankey diagram depicting the mapping of ground truth classes to the top 5 most common diagnostic entities in the test set in each class (left). Malignant melanoma was not in the top 5 but included here due to its clinical importance. Also shown is the proportion of images correctly classified, along with the distribution of misclassifications and unclassified specimens (those for which confidence score was below the threshold) at confidence Level 1 (right). The width of each bar is proportional to the corresponding number of specimens found in the 3-lab test set.}
     \label{fig:sankey}
 \end{figure}

To achieve high classification accuracy in the presence of a wide range of variability in tissue appearance between labs, a unique calibration set (about 520 WSIs) was collected from each lab and used to fine-tune the final classifier (CNN-3). Results are reported only on the test set, consisting of 13,537 WSIs from the 3 test labs which were not used in model training or development. The deep learning system effectively classifies WSIs into the 4 classes with an overall accuracy of 78\% before thresholding on confidence score.
Importantly, in specimens whose predictions exceeded the confidence threshold, the PDLS achieved an accuracy of 83\%, 94\%, and 98\% for confidence levels 1, 2 and 3, respectively. 
Performance of the PDLS is characterized with receiver operating characteristic (ROC) curves, shown for each of the 4 classes in Fig.~\ref{fig:ROC}a-d at each confidence level; as confidence level increases, a larger percentage of images do not meet the threshold and are excluded from the analysis, as indicated by the colorbar. 
At Levels 1, 2, and 3, the percentage of test specimens exceeding the  confidence threshold was 83\%, 46\% and 20\%, respectively.
Area under the curve (AUC) increased with increasing confidence level. Similar results are shown for Level 1 for each test lab in Fig.~\ref{fig:ROC}f-i which compare AUC and percentage of specimens confidently classified between the 3 labs. 
Fig. \ref{fig:sankey} shows the mapping of ground truth class to the proportion correctly predicted as well as proportions confused for each of the other classes or remaining unclassified (at Level 1) due to lack of a confident prediction or absence of any ROI detected by CNN-2.  Additionally, this figure shows the most common ground-truth diagnoses in each of the 4 classes found in the test set.

\subsection{Reduction of Inter-site Variance}
\begin{figure}
     \centering
     \includegraphics[width = \linewidth]{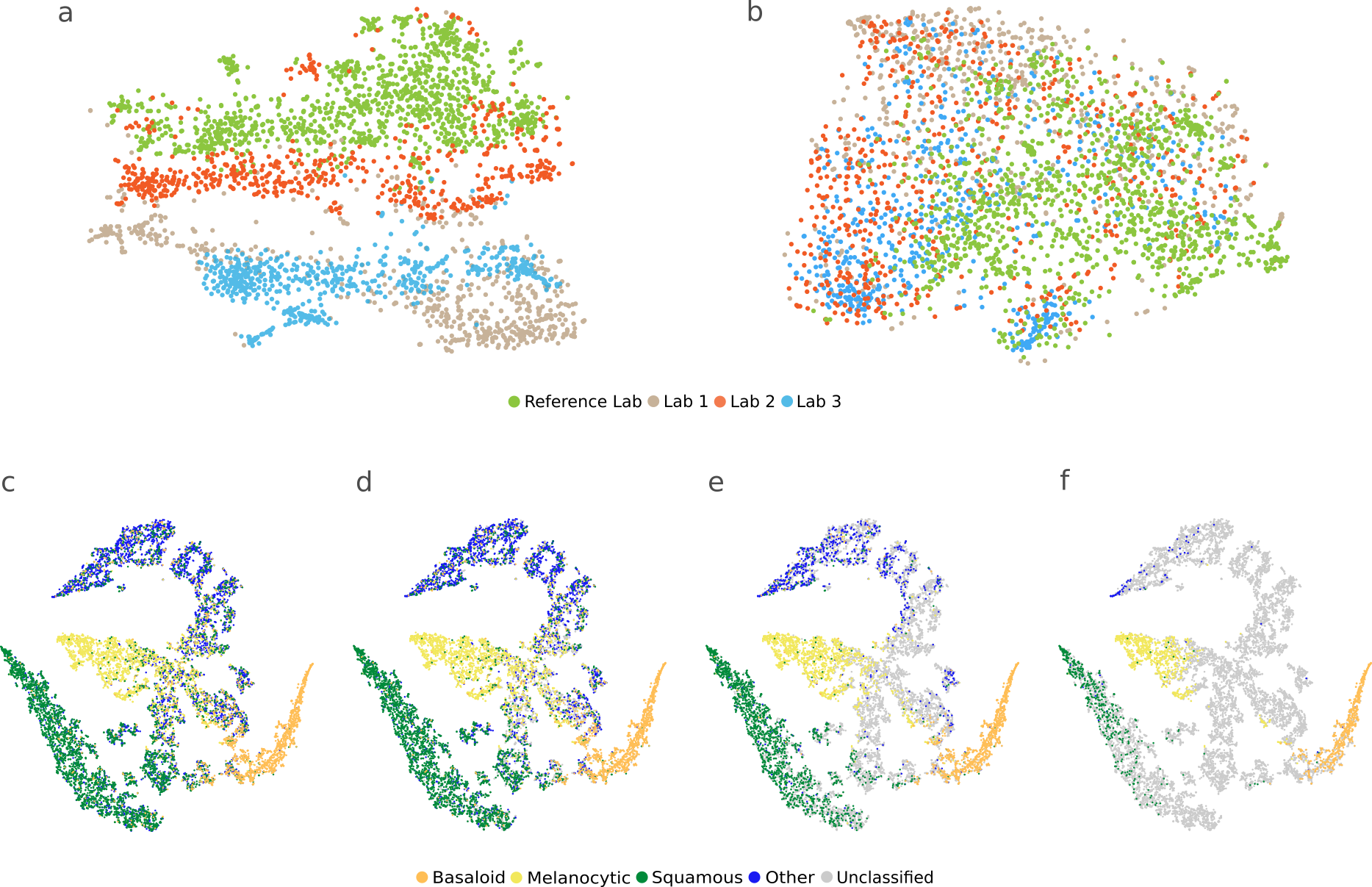}
     \caption{Image feature vectors are shown in 2-dimensional t-distributed stochastic neighbor embedded (t-SNE) plots. \emph{Top:} Feature embeddings from CNN-2 are shown with a) no prior image adaptation and b) when image adaptation (using CNN-1) is performed prior to performing region of interest (ROI) extraction using CNN-2. Each point is an image patch within a whole slide image (WSI), colored by lab. \emph{Bottom:} Feature embeddings from CNN-3, where each point represents a specimen and is colored according to ground-truth classification. All specimens are classified at baseline (a), where \emph{(d-f)} show increasing confidence thresholds (d=Level 1, e=Level 2, f=Level 3), with specimens not meeting the threshold in gray}. 
     \label{fig:t-SNE}
 \end{figure}

To demonstrate that the image adaptation performed by CNN-1 effectively reduces inter-site variations, we used t-distributed stochastic neighbour embedding (t-SNE) to compare the feature space of CNN-2 with and without first performing the image adaptation step. We show CNN-2's embedded feature space \emph{without} first performing image adaptation in Fig.~\ref{fig:t-SNE}a; Fig.~\ref{fig:t-SNE}b
then shows the embedded feature space from CNN-2 when image adaptation is performed first. Inclusion of the image adaptation step results in more overlapped distributions in feature space than those produced without using image adaptation; this transformation into a common feature space allows the system to perform high-quality classification regardless of staining technique or scanner used.

\subsection{Effective Class Separation}
Additionally, we used t-SNE to show class separation based on the internal feature representation learned by the final classifier (CNN-3), as shown in Fig.~\ref{fig:t-SNE}c.
Each point in these t-SNE plots represents a single specimen with color denoting its ground-truth class. Figs.~\ref{fig:t-SNE}d-f show the same information when thresholding at each of the 3 confidence levels (1-3, respectively), indicating in gray the specimens left unclassified at each. The clustering shows strong class separation between the 4 classes, with stronger separation and fewer specimens classified as confidence level increases.

\subsection{Timing Profile}
\begin{figure}
     \centering
     \includegraphics[width = \linewidth]{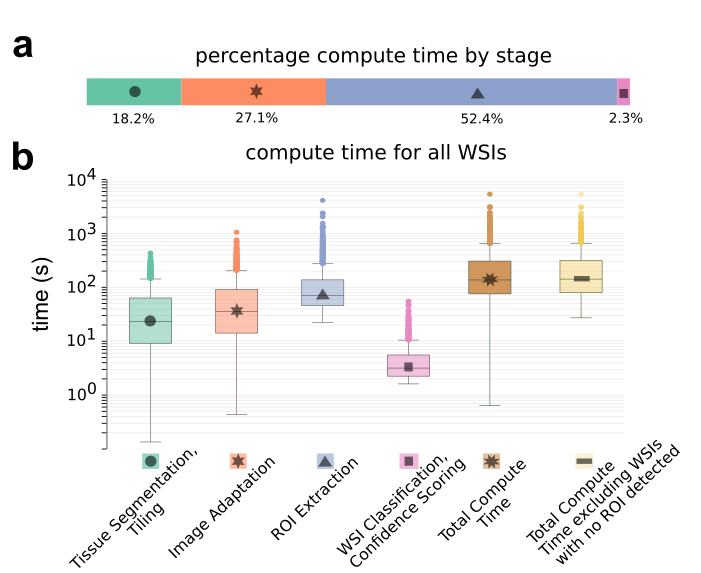}
     \caption{\textbf{PDLS compute time for whole slide images on the calibration sets from the 3 test labs.} (a) The median percentage of total computation time for each stage in PDLS is shown. (b) A boxplot of the computation time in seconds required at each stage of the pipeline is shown on a logarithmic scale, along with total end-to-end execution time for all images (dark brown, median 137s), and excluding images for which no regions of interest are detected (light brown, median 142s).}
     \label{fig:timing}
 \end{figure}

It is important that execution time for any system intended to be implemented in a lab workflow be low enough to not present a bottleneck to diagnosis. Therefore, the proposed system was designed to be parallelizable across WSIs to enhance throughput and meet the efficiency demands of the real-world system. On a single compute node (described in Methods), the median processing time per WSI was 137 seconds, with overall throughput of 40 WSIs/hour. Fig.~\ref{fig:timing}a shows the median time consumed by each stage in the pipeline, and Fig.~\ref{fig:timing}b shows box-plots of time at each stage, as well as end-to-end execution time. 

\section{Discussion}
Our work demonstrates the ability of a multi-site generalizable PDLS to accurately classify the majority of specimens in a routine dermatopathology lab workflow.

Developing a deep-learning-based classification which translates across image sets from multiple labs is non-trivial.\cite{Tellez2019,campanella,Ciompi2017} Without compensation for image variations, non-morphological differences between data from different labs are more prominent in the feature space than morphological differences between the specimens ultimately belonging to the same diagnostic classification. This is demonstrated in Fig. \ref{fig:t-SNE}a, in which the image patches cluster according to the lab that prepared and scanned the corresponding slide. When image adaptation is performed prior to computing image features, the images do not strongly cluster by lab (Fig. \ref{fig:t-SNE}b).
In this study, we demonstrate that a PDLS trained on a single Reference Lab can be effectively calibrated to 3 additional lab sites. Figs. \ref{fig:t-SNE}c-f show strong class separation between 4 classes, and this class separation strengthens with increasing confidence threshold. Intuitively, low-confidence images cluster at the intersection of the 4 classes. Strong class separation is reflected also in the ROC curves, which show high AUC across classes and labs, as seen in Fig. \ref{fig:ROC}. AUC increases with increased confidence level, demonstrating the utility of confidence score thresholding as a tunable method for excluding poor model predictions. Figs. \ref{fig:ROC}d shows relatively worse performance in the Other class. In \ref{fig:t-SNE}c it can be seen that there is some overlap between the Squamous and Other classes in feature space; Fig. \ref{fig:sankey} also shows some confusion between these two classes, but overall, demonstrates accurate classification of the majority of specimens from each class.

The majority of previous deep learning systems in digital pathology have been validated only on a single lab or scanner's images,\cite{Olsen, campanella,Li2019} curated datasets that ignored a portion of lab volume within a speciality,\cite{campanella,Olsen,Ghaznavi2013} and tested on small and unrepresentative datasets,\cite{Olsen,Hart,Ghaznavi2013,Li2019} excluded images with artifacts\cite{Olsen,campanella,Bulten2019} or selectively reverse image "ground truth" retrospectively for misclassifications\cite{campanella} and train patch- or segmentation-based models while using traditional computer vision or heuristics to arrive at a whole slide prediction.\cite{Olsen,breastcamelyon,Bulten2019} These methods do not lend themselves to real-world enabled deep learning systems that are capable of operating independent of the pathologist and prior to pathologist review. These systems would require some human intervention before they can provide useful information about a slide, and therefore do not enable improvements in lab workflow efficiencies.

In contrast, our PDLS is trained on \emph{all} available slides-- images with artifacts, 
slides without tissue on them, slides with poor staining or tissue preparation, slides exhibiting rare pathology, and those with very subtle evidence of pathology. All of this variability in the data necessitates that our PDLS is capable of determining when it is not likely to make a well-informed prediction. This is accomplished with a confidence score, which can be thresholded to obtain better system performance as shown in Fig.~\ref{fig:ROC}a-e.
Correlation between system accuracy and confidence was established \emph{a priori} using only the Reference Lab validation set (Fig. \ref{fig:ROC}e) to fix the 3 confidence thresholds. By fixing thresholds \emph{a priori} we establish that they are generalizable. Campanella et al.\cite{campanella} have attempted to similarly set a classification threshold which yields optimal performance; however, they perform this thresholding using the last layer output of a model, on the same test set in which they report it yielding 100\% sensitivity; therefore they do not demonstrate the generalizability of this tuned parameter. Secondly, as Gal et. al\cite{gal2016dropout} demonstrate, a model's predictive probability (last layer output) cannot be interpreted as a measure of confidence.

\par We report all performance measures (accuracy, AUC) at the level of a specimen, which may consist of several slides, since diagnosis is not reported at the slide level in dermatopathology. We aggregate all slide-level decisions to the specimen level as reported in Methods; this is particularly important as not all slides within a specimen will exhibit pathology, and therefore an incorrect prediction can be made if slide-level-reporting is performed. Similar systems \cite{Olsen,Hart,Ing2018,campanella} have not attempted to solve the problem of aggregating slide-decisions to the specimen level at which diagnosis is performed.
\par For the PDLS to operate before pathologist assessment, the entire pipeline must be able to run in a time period that avoids delaying the presentation of a case to the pathologist. The compute time profile shown in Fig. \ref{fig:timing}a-b demonstrates that the PDLS can classify a WSI in under 3 minutes in the majority of cases, which is on the same order of the amount of time it takes for today's scanners to scan a single slide. There was considerable variation in this number due to  a large amount of variability in the size of the tissue. However, it is important to note that this process can be infinitely parallelized across WSIs to enhance throughput. Additional optimization of this process is possible and is the subject of future work. 
There are several limitations to the current PDLS which are shared by previous implementations of deep learning image classification in digital pathology. First, when diagnosing a specimen, pathologists often have access to additional clinical information about the case, whereas our PDLS uses only WSIs to make a prediction. Training the PDLS with this additional clinical context as input would likely improve accuracy in some cases. A second limitation is that all existing systems for pathology classification attempt to put restrictions on the biology, namely that a WSI or a specimen can only represent a single diagnosis. Rarely (2-3\% of specimens), a specimen should be labelled with more than one class. We did not train the current PDLS to handle this special case since the available sample of images with dual ground-truth class is small; however, this will be a subject of future research.
\par While the current PDLS does not make diagnostic predictions, its classification has the potential to increase diagnostic efficiency and consistency in several scenarios. For example, pathologists might choose to prioritize certain classes, e.g. Melanocytic, that may contain more difficult cases, requiring longer review time, additional levels ordered, or ancillary testing such as immunostains. Similarly, a dermatologist who interprets biopsies could choose to only receive cases classified as Basaloid, and avoid receiving many inflammatory cases or melanocytic lesions which might be sent for referral. The tunability of the confidence threshold in the model as a near-final step in assigning a classification has further implications for how this deep learning system might be utilized in practice. For applications that depend on high-sensitivity classification (e.g.  treating classification as a form of quality assurance to assist in avoiding missed diagnosis of melanomas, which should exist in the Melanocytic classification), a higher confidence threshold might be set. Similarly, for an application that depends less on specificity (e.g. triage of cases to balance pathologists' workloads) the desired confidence threshold could be lower, thereby avoiding an overly-large set of unclassified specimens. Finally, as hierarchical classification models have been shown to outperform flat classifiers,\cite{Silva-palacios2018} we expect that the current PDLS serves as a basis for extension to diagnostic classification systems. This would enable further prioritization of more critical cases, such as those presenting features of melanoma.

\subsection{Conclusion}
The techniques presented herein--namely deep learning of heterogeneously-composed classes, and confidence-based prediction screening-- are not limited to application in dermatopathology or even pathology, but broadly demonstrate potentially effective strategies for translational application of deep learning in medical imaging. The PDLS presented delivers accurate prediction, regardless of scanner type or lab, and requires minimal calibration to achieve accurate results for a new lab. The system is capable of assessing which of its decisions are viable based on a computed confidence score, and thereby can filter out predictions that are unlikely to be correct. This confidence-based strategy is broadly applicable for achieving the low error rates necessary for the practical use of machine learning in challenging and nuanced domains of medical disciplines.

\section{Methods}
\subsection{Data Used in Development}
The proposed system was developed in its entirety using H\&E-stained WSIs from Dermatopathology Laboratory of Central States, which is referred to as the Reference Lab in this work. All slides from this Reference Lab were scanned using the Leica Aperio AT2 Scanscope (Aperio, Leica Biosystems, Vista, California). This dataset is made up of two subsets, the first (3,070 WSIs) consisting of images representing commonly diagnosed dermatopathologic entities, and the second (2,000 slides) consisting of all cases accessioned during a discrete period of time, representing the typical distribution seen by the lab. This combined Reference Lab set of 5,070 WSIs was partitioned randomly into training~(70\%), validation~(15\%), and testing~(15\%) sets, such that WSIs from any given specimen were not split between sets.

\subsection{Taxonomy}
The design of target classes in this study is heavily influenced by the prevalence of each class's constituent pathologies and the presence of visually- and histologically-similar class-representative features. They capture, in roughly equal proportion, the majority of diagnostic entities seen in a dermatopathology lab practice. Specifically, we perform classification of WSIs into four classes: Basaloid, Squamous, Melanocytic, and Others. These four classes are defined by the following histological descriptions of their features:
\begin{enumerate}
    \item \emph{Basaloid}:  
Abnormal proliferations of basaloid-oval cells having scant cytoplasm and focal hyperchromasia of nuclei; cells in islands of variable size with round, broad-based and angular morphologies; peripheral palisading of nuclei, peritumoral clefting, and a fibromyxoid stroma.
    \item \emph{Squamous}:  
Squamoid epithelial proliferations ranging from a hyperplastic, papillomatous and thickened spinous layer to focal and full thickness atypia of the spinous zone as well as invasive strands of atypical epithelium extending into the dermis at various levels.     
    \item \emph{Melanocytic}: 
Cells of melanocytic origin in the dermis, in symmetric, nested, and diffuse aggregates and within the intraepidermal compartment as single cell melanocytes and nests of melanocytes. Nests may be variable in size, irregularly spaced, and single cell melanocytes may be solitary, confluent, hyperchromatic, pagetoid and with pagetoid spread into the epidermis. Cellular atypia can range from none to striking anaplasia and may be in situ or invasive.  
    \item \emph{Other}: 
Morphologic and histologic patterns that include either the absence of a specific abnormality or one of a wide variety of other neoplastic and inflammatory disorders which are both epithelial and dermal in location and etiology, and which are confidently classified as not belonging to Classes 1-3.
\end{enumerate} These four classes account for more than 200 diagnostic entities in our test set, and their mapping to the most prevalent diagnostic entities in the test set is illustrated in Fig.~\ref{fig:sankey}.

 \subsection{System Design and Training}
Our image processing pipeline for the PDLS is illustrated in Fig. \ref{fig:processFig}.
The PDLS takes as input a WSI, segments out regions containing tissue, and divides these regions into a set of tiles, each of size 128 $\times$ 128 pixels. The process of assigning a label to a WSI using this set of tiles is comprised of three stages: 1) Image Adaptation, 2) Region of Interest Extraction, and 3) WSI Classification.

Since the PDLS is trained on only a single lab's data, it is critical to perform image adaptation to adapt images received from test labs to a domain in which the image features are interpretable by the PDLS. Without adaptation, unaccounted-for variations in the images due to staining and scanning protocols can critically affect the performance of CNNs.\cite{Tellez2019, campanella,Ciompi2017} The PDLS performs image adaptation using a CNN (referred to as CNN-1), which takes as input an image tile and outputs an adapted tile of the same size and shape but with standardized image appearance. CNN-1 was trained using 300,000 tiles from the Reference Lab and mimics the average image appearance from the Reference Lab given an input tile.

Subsequently, ROI extraction is performed using a second CNN (referred to as CNN-2). This CNN is trained using expert annotations by a dermatopathologist as the ground truth. It performs a segmentation of regions exhibiting abnormal features indicative of pathology. The model takes input of a single tile and outputs a segmentation map. Tiles are selected corresponding to the positive regions of the segmentation map; the set of all identified tiles of interest, $t$ from a WSI is passed on to the final stage classifier.

The final WSI classification is then performed using a third CNN (CNN-3), which predicts a label, $l$ for the set of tiles $t$ identified by CNN-2 where: 
\begin{equation}
l \in \{\rm Basaloid,~Squamous,~Melanocytic,~Others\} .
\end{equation}
CNN-3 additionally outputs a confidence score for each WSI. In clinical practice, and in our dataset, diagnostic labels are reported at the level of a specimen, which may be represented by one or several WSIs. Therefore, the predictions of the PDLS are aggregated across WSIs to the specimen level; this is accomplished by assigning to a given specimen the maximum-confidence prediction across all WSIs representing that specimen.

\subsection{Calibration and Validation for Additional Sites}
To demonstrate its robustness to variations in scanners, staining, and image acquisition protocols, the PDLS was tested on 13,537 WSIs collected from 3 dermatopathology labs, representing two leading dermatopathology labs in top academic medical centers (Dermatopathology Center at Thomas Jefferson University and the Department of Dermatology at University of Florida College of Medicine) and a high volume national private dermatopathology laboratory (Cockerell Dermatopathology).We refer to these as \emph{test labs}. Prior to the study, each lab sought study approval from the appropriate Institutional Review Board and was exempted. Each lab performed scanner validation prior to data collection, according to the guidelines of the College of American Pathologists.\cite{Pantanowitz2013} Each test lab selected a date range within the past 4 years (based on slide availability)
from which to scan a sequentially accessioned set of approximately 5,000 slides. Each of the 3 test labs scanned their slides using a different scanner vendor and model. Scanner models used were: Leica Aperio AT2 Scanscope Console (Leica Biosystems, Vista, California), Hamamatsu Nanozoomer-XR (Hamamatsu Photonics, Hamamatsu City, Shizuoka, Japan), and 3DHistech Pannoramic 250 Flash III (3DHistech, Budapest, Hungary). All parameters and stages of the PDLS pipeline were held fixed after development on the Reference Lab, with the exception of CNN-3, whose weights were fine-tuned independently for each lab using a calibration set of approximately 520 WSIs. (We refer to this process as calibration). The calibration set for each lab consisted of approximately 500 sequentially-accessioned WSIs (pre-dating the test set) supplemented by 20 additional WSIs from melanoma specimens. Of these calibration images, 80\% were used for fine-tuning, and 20\% for lab-specific validation of the fine-tuning and image adaptation procedures. Specimens from the same patient were not split between fine-tuning, validation and test sets. After this calibration, all parameters were permanently held fixed, and the system was run only once on each lab's test set of approximately 4,500 WSIs (range 4451 to 4585)-- 13,537 in total.
\subsection{Confidence Scoring and Threshold Computation}
\label{sec:conf}
Gal et al.\cite{gal2016dropout} propose a method to reliably measure the uncertainty of a decision made by a classifier. We have adapted this method for confidence scoring of the decision made by PDLS. To determine a confidence score for a WSI we perfom prediction on the same WSI repeatedly for (using CNN-3) several times by omitting a random subset of neurons (here 70\%) in CNN-3 from the prediction. Each repetition results in a prediction made using a different subset of feature representations. 
Here, we use $T=30$ repetitions, where each repetition $i$ yields a prediction $P_i$, a vector of sigmoid values of length equal to the number of classes. Each element of $P_i$ represents the binary probability, $p_{i,c}$, of the corresponding WSI belonging to class $c$. 
The confidence score $s$ for a given WSI is then computed as follows:
\begin{equation}
    s = \max_{c}\left({\frac{\sum_{i=1}^{T} p_{i,c} }{T}}\right)
\end{equation}
The class associated with the highest confidence $s$ is the predicted class for the WSI. Finally, the specimen prediction is assigned as the maximum-confidence prediction of its constituent WSI predictions.
If a specimen's confidence score is below a certain threshold, then the prediction is considered unreliable and the specimen remains unclassified. 
Three threshold values for the confidence score were selected for analysis; these were determined during the development phase, using only the Reference lab's data, because this confidence threshold is a parameter which can tune model performance. Confidence thresholds were selected such that discarding specimens with sigmoid confidence lower than the threshold yielded a pre-defined level of accuracy in the remaining specimens of the validation set of the Reference Lab. The three target accuracy levels were 90\%, 95\% and 98\%; the corresponding sigmoid confidence thresholds of 0.33, 0.76, and 0.99 correspond to confidence Levels 1, 2, and 3 respectively; these confidence thresholds were held fixed, and applied without modification to the test sets from the 3 test labs.
\subsection{Compute Time} 
Compute time profiling of the PDLS was performed on an Amazon Web Services EC2 P3.8x large instance equipped with 32 core Intel Xeon E5-2686 processors, 244 GB RAM, and four 16GB NVIDIA Tesla v100 GPUs supported by NVLink for peer-to-peer GPU communication. Compute time was measured on the calibration sets of each of the the test labs.

\begin{addendum}
\item
We would like to thank Hamamatsu and Epredia for loaning whole slide image scanners. We thank Katherine Tesno, Denise Lunsford, Cindy Jones, Cassandra Morgan, Valerie Matteo, Doa Salabi, and Craig Reed all for their hard work in scanner operation and data collection, and Mary Bohannon for study coordination, IRB support and scanner operation. We are grateful also to Michael Kent, Ph.D. for scientific advice and discussion, Nathan Buchbinder for help with study coordination and manuscript review, Saul Kohn, Ph.D. for manuscript review, and Addie Walker, M.D. and Vladimir Vincek, M.D., Ph.D. for review of specimens.

\end{addendum}
\section*{References}
\bibliography{references}
\end{document}